\documentclass[letterpaper, 10pt, conference, leqno]{ieeeconf}
\IEEEoverridecommandlockouts
\usepackage[english]{babel}
\usepackage{upgreek}
\usepackage{amsfonts}
\usepackage{multicol}
\usepackage{amssymb}
\usepackage{amsmath}
\usepackage{cite}
\usepackage{graphicx}
\usepackage{hyperref}

\let\labelindent\relax
\usepackage{enumitem}
\usepackage[export]{adjustbox}
\usepackage{mathtools}

\usepackage{xcolor}

\title{\LARGE \bf Path Integral Bottleneck: \\ 
An Algorithm-Agnostic Framework of Computation and Control}
\author{Justin Ting, Jing Shuang (Lisa) Li
\thanks{J.T. and J.S.L. are with the Department of Electrical Engineering and Computer Science at the University of Michigan, Ann Arbor. {\tt\small sigfyg@umich.edu, jslisali@umich.edu}.}
}

\begin{document}

\maketitle

\begin{abstract}
Executing a control sequence requires computation. While this is a simple observation, developing a framework that relates a controller's required computation to its ability to successfully control a system (e.g. lower control cost) is challenging, especially when the controller appears on alternative compute platforms (e.g. biological neural networks). More specifically, we want a framework where, given an observed closed-loop trajectory, we can quantify the computation effort needed to produce that trajectory. To enable effective comparisons of closed-loop systems across alternative compute platforms, we present the Path Integral Bottleneck (PI-IB), a method to produce an analytical, algorithm-agnostic description of the compute-control relationship. With the PI-IB framework, we can plot tradeoffs between performance and computation effort for any given plant description and control cost function. Simulations of the cart-pole reveal fundamental control-compute tradeoffs, exposing regions where the task performance-per-compute is higher than others.

% quantify how sophisticated a feedback controller is by only observing behavior

% Intuitively, a high-effort, fine-grained computation should result in better control (e.g. lower control cost), whereas little to no computation effort would lead to worse control. Quantifying the tradeoff between control performance and compute effort, unbounded by implementation details, would allow effective comparisons of closed-loop systems across alternative compute platforms (e.g. biological neural networks). To quantify this tradeoff, we present the Path Integral Bottleneck (PI-IB), a method to produce an analytical, algorithm-agnostic description of the compute-control tradeoff. With the PI-IB framework, we can plot tradeoffs between performance and computation effort for any given plant description and control cost function. Simulations of the cart-pole reveal fundamental control-compute tradeoffs, exposing regions where the task performance-per-compute is higher than others.

% The the Path Integral (PI) optimal control and Information Bottleneck (IB) frameworks.
% Both frameworks provide flexible and probabilistic descriptions of control. 
% The PI does not limit itself to a particular control law, and the IB is not bound to any specific state encoding. 
% We provide PI-IB formulations for both continuous and discrete random variables. 
% \textit{Keywords-} biology, optimal control, robotics
\end{abstract} 

\section{Introduction}
% quantify how sophisticated a feedback controller is by only observing behavior
Evaluating a control algorithm's complexity and resource requirements typically assumes execution on a microprocessor, where we know the implementation details. However, when control is executed on alternative compute platforms (e.g. biological neural network), implementation details are not available, making comparisons to microprocessor implementations difficult. A proper complexity characterization that can account for this limitation would facilitate an integrated and improved understanding of control algorithms in nature (e.g. animal movement). In this work, we aim to produce such a characterization over alternative compute implementations by quantifying a tradeoff between compute effort and controller performance.

Inspired by Kolmogorov complexity\cite{kolmogorov_vityani}, we quantify the complexity of an observed trajectory based on its potential to be random. Kolmogorov randomness connects the probability of a computation's results (e.g. the probability random inputs can produce Shakespeare) to computation effort. In the controls context, we can think about the probability that random signals successfully control a system. To illustrate, compare Bang-Bang to MPC. At any given time, there is a 50\% chance of guessing the control signal correctly for Bang-Bang, so it is more predictable compared to MPC. Intuitively, Bang-Bang also requires less computation effort than MPC. Therefore, we can associate Bang-Bang's predictability with a lower compute cost. 
% Bang-Bang intuitively requires less compute effort. One would not use MPC for an application where Bang-Bang suffices, but for a significantly complex task, the additional computation effort of MPC is justified. 

\begin{figure}
\includegraphics[scale=.51]{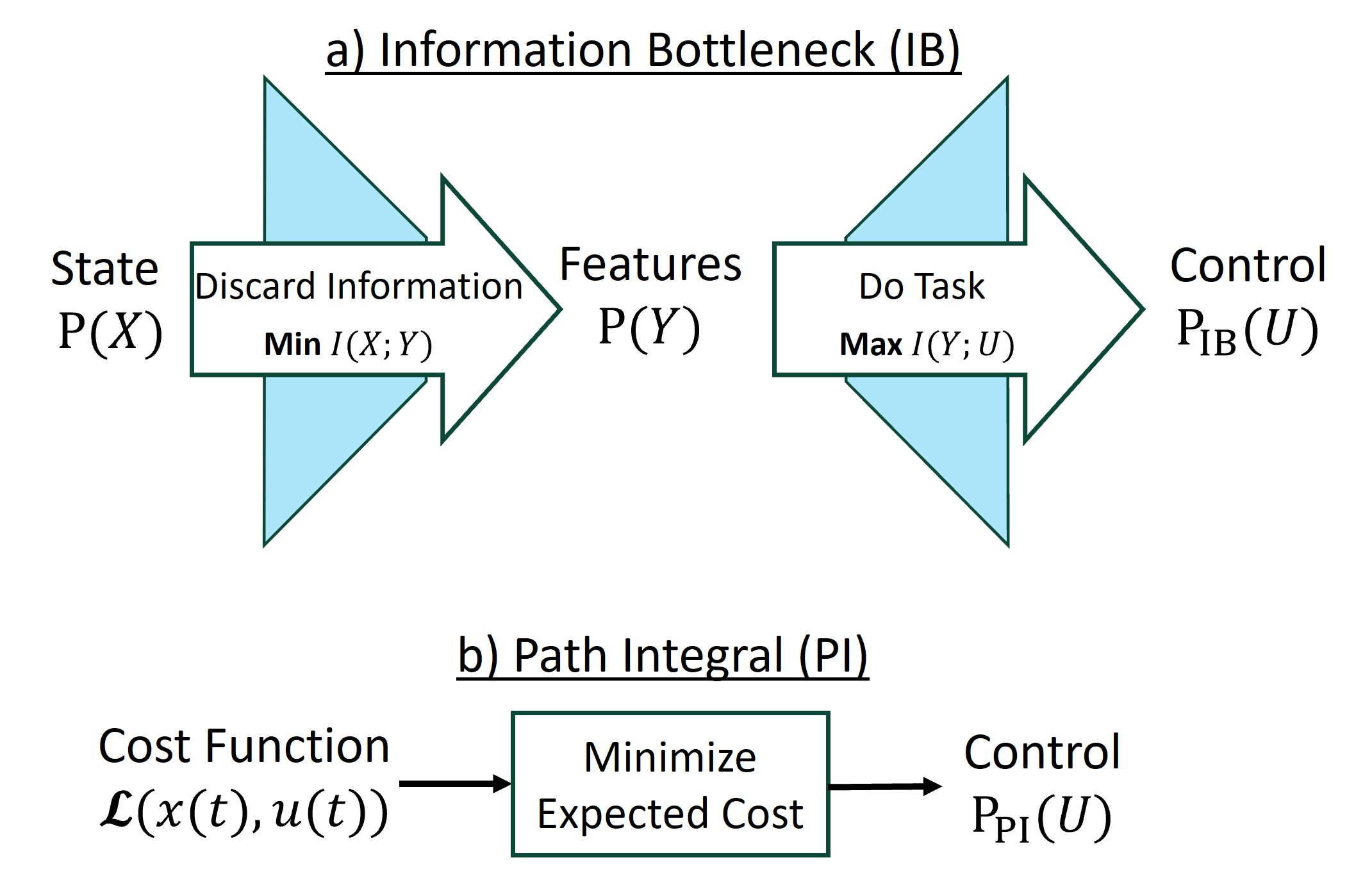}
\caption{Two probability distributions over control sequences: $P_{IB}(U)$ through the Information Bottleneck and $P_{PI}(U)$ through the Path Integral. Neither impose an interpretation on the probabilities. a) $P_{IB}(U)$ is constrained by the bottleneck size, which is the mutual information $I(X;Y)$. $X$ is not necessarily a state space vector (e.g. images). b) $P_{PI}(U)$ is constrained by a cost function $\mathcal{L}(x(t), u(t))$. Here, $x$ is a state space vector.}
\label{fig:two_ways}
\vspace{-0.7cm}
\end{figure}

Two existing frameworks that can evaluate the randomness of a trajectory are the Information Bottleneck (IB) and Path Integral (PI) control (Fig.~\ref{fig:two_ways}). The IB method optimizes the trade-off between information loss and task performance \cite{tishby2000informationbottleneckmethod}\cite{NIPS2003_gaussian_bottleneck}. For example, when labeling images, minimize the number of features extracted while labeling accurately. To investigate the IB in the controls context, we combine it with PI, an optimal control framework that statistically minimizes a running cost by generating random samples of control signals\cite{theodorouGeneralizedPathIntegral}\cite{williamsInformationTheoreticMPC2017}. 

The relationship between the IB and control, such as the case of restricted communication or state estimation, has been explored in various ways. For a given communication channel between an estimator and LQG controller, one can combine the LQG objective with a communication cost to solve for an optimal code length and rate in addition to an optimal control input \cite{rate_vs_cost_2017}\cite{control_comm_constraints_2004}\cite{LQG_comm_constrainted_1998}\cite{Borkar1997}. For works that explicitly use an IB, multiple problems can be formulated. For example, one can find a reduced-dimension system realization by compressing a control sequence into a representation that best predicts an output sequence \cite{Tishby_past_future_bottleneck_2015}. The IB can also be used to explore the tradeoff between a controller's memory of past states and its performance \cite{Tishby_min_info_LQG_2016}. When deep learning is used to measure states from data, the IB can be used to determine which features are most relevant to the control task \cite{separation_principle_bottleneck_2017}. More recently, a bottleneck approach has been used to explore stochastic policies, for both state estimation \cite{task_driven_estimation_control_ICRA_2018} and extracting states from data \cite{task_driven_control_bottleneck_RSS_2020}. 

Previous works focus on linear systems, the role of state history in making control decisions, and interpret randomness as noise. However, they do not explore a fully nonlinear system implementation, address the role of planning in feedback, or provide a complexity interpretation. In this work, we introduce the Path Integral Bottleneck (PI-IB) framework to analyze the complexity of a feedback controller producing a sophisticated trajectory, while extending to nonlinear systems. PI-IB combines the PI\cite{williamsInformationTheoreticMPC2017} and IB\cite{tishby2000informationbottleneckmethod}, two descriptions of a control signal's randomness. These two approaches can be leveraged to provide a way to understand control sequences as computationally constrained processes. 

The PI does not impose a specific way of minimizing the cost function. In other words, we are not bound to any particular control law, freeing us from interpretations based on policy parameters. Furthermore, the PI admits unconventional models and cost functions (e.g. hybrid models and lookup-table cost functions). The IB is also flexible in that it does not impose a specific encoding an algorithm could use. The novel combination of IB and PI allows an \textbf{encoding and control-law agnostic} exploration of the compute-control tradeoff, allowing us to evaluate algorithms across alternative compute platforms without being constrained by implementation details. To demonstrate our framework, we simulate and plot the fundamental compute-control tradeoffs for the cart-pole problem, sweeping over a range of behaviors (Section \ref{sec:simmy}). The PI-IB analysis reveals that balancing and swing-up have distinct compute-control tradeoffs.

\textbf{Notation:} For a random variable $X$, its distribution is $P(X)$ for both discrete (mass) and continuous (density) cases. All sums over discrete variables imply its integral counterpart for continuous variables. The probability mass/density for a realization $x$ is $p(x) = P(X=x)$. For conditional distributions, $P(X|y) = P(X|Y=y)$ and $p(x|y) = P(X=x|Y=y)$. All logarithms are base $e$.

\section{Background}
We separately review the technical background of PI and IB required for the PI-IB. The IB searches over probability distributions to balance compression and performance; the PI defines a probability distribution over trajectories based on a running cost (Fig. \ref{fig:two_ways}). 

% $H_P(X)$ is the entropy of the distribution $P(X)$, and $D_{KL}(P(X)||Q(X))$ is KL-divergence between the distributions $P(X)$ and $Q(X)$.

\subsection{Information Bottleneck (IB) Method}\label{subsec:infoneck}
\begin{figure}
\includegraphics[scale=0.8, trim=0cm .5cm 0cm 0cm, center]{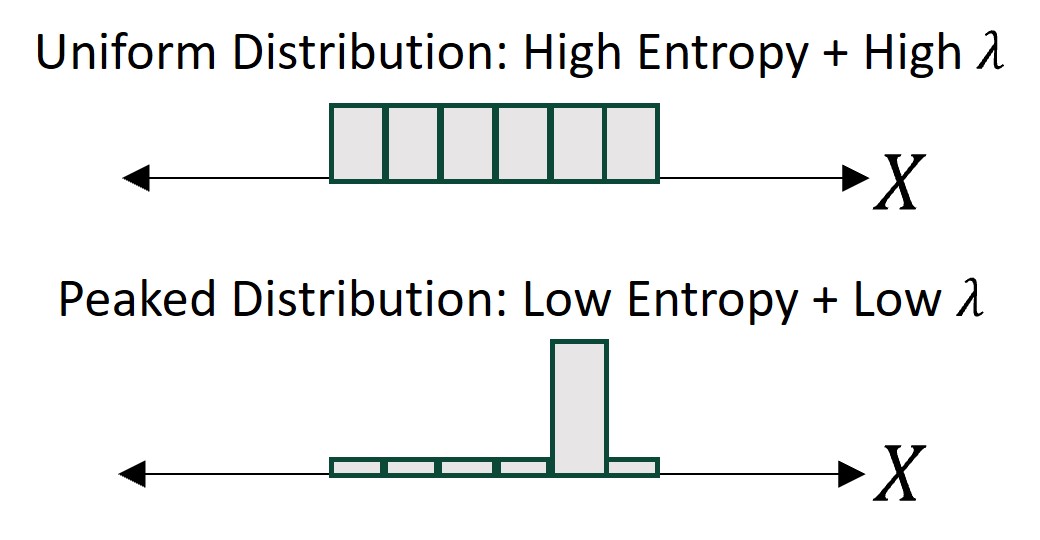}
\caption{A distribution $P(X)$'s shape changes its entropy and the random variable's predictability. The parameter $\lambda > 0$ from the PI formulation changes how flat or peaked the distribution is.}
\label{fig:simple_entropy_example}
\vspace{-0.7cm}
\end{figure}
Before directly addressing the IB, we will go over notions of information and entropy that guide our PI-IB framework. A probability distribution $P(X)$ encodes predictability of the random variable $X$. A flat, uniform distribution makes it difficult to predict $X$'s outcome, but a peaked, dirac delta distribution makes it easy to predict $X$. This is quantified with entropy, calculated as below:
\begin{equation}
\text{Entropy:} \hspace{0.5cm} H_P(X) \triangleq -\sum_X p(x)\log p(x)
\end{equation}
The unit for entropy is bits (base-2) or nats (base-$e$). We use nats in this work. For discrete random variables, the entropy $H_P(X)$ is maximized when $P(X)$ is a uniform distribution, and minimized with dirac delta distribution. In other words, a variable with high entropy is difficult to predict (Fig. \ref{fig:simple_entropy_example}).

Mutual information between two variables $I(X;Y)$ is a difference of entropies, calculated as below, where $H_P(X|Y)$ is conditional entropy. 
\begin{equation}
\begin{aligned}
I(X;Y) &\triangleq H_P(Y) - H_P(Y|X) \\
&= H_P(X) - H_P(X|Y) \geq 0 \\[0.2cm]
H_P(Y|X) &\triangleq -\sum_{X,Y} p(x,y)\log p(y|x)
\end{aligned}
\end{equation}
$H_P(Y)$ is $Y$'s predictability \textit{without} knowledge of $X$, and $H_P(Y|X)$ is $Y$'s predictability \textit{with} knowledge of $X$. Therefore, $I(X;Y)$ quantifies the knowledge of $Y$ gained after acquiring knowledge of $X$ (and vice-versa).

The information bottleneck (IB) method is concerned with mutual information in the Markov dependency relationship $X\rightarrow Y\rightarrow U$\footnote{Our variable choice deviates from the convention $X\rightarrow (T \ \text{or} \ \tilde{X}) \rightarrow Y$ to fit the controls context.}, where $Y$ is a representation of $X$ used to predict $U$. These three random variables can accept a wide range of interpretations. 
% For example, when labeling images, $X$ represents images, $Y$ represents extracted features, and $U$ represents labels.  
In the control theory context, $X$ can represent a history of states, fused sensors, images, etc. For the rest of this work, $X$ is simply a state-space vector, although our PI-IB framework can admit other interpretations. $Y$ can represent an observer's output, a state estimate, an arbitrary encoding, partitions or clusters of $X$, extracted features, etc. $U$ is the control signal. We address interpretations of $U$'s entropy in Section \ref{sec:PIIB}.

The IB method \cite{tishby2000informationbottleneckmethod}\cite{nonlinear_bottleneck} presents the following optimization problem.
\begin{equation}
\min_{P(Y|X)} \quad  \mathcal{F}_{IB} = I(X;Y) \hspace{1cm} \text{s.t.} \quad  I(Y;U) \geq R
\end{equation}
Minimizing $I(X;Y)$ equates to shrinking the information bottleneck (discarding information from $X$), while the constraint $I(Y;U) \geq R$ preserves the performance and predictability of $U$ (Fig. \ref{fig:two_ways}). Shrinking the bottleneck reduces performance, so there is a tradeoff. In the controls context, minimizing this cost function is equivalent to minimizing the number of features ($Y$) an algorithm should extract from all available information ($X$) in order to produce an acceptable control signal ($U$). The constraint $I(Y;U) \geq R$ introduces practical difficulties when finding a minimum, so typically an equivalent problem is formulated as below: 
\begin{equation}\label{eqn:F_IB}
\min_{\substack{P(Y|X),\\ P(Y), \\ P(U|Y)}} \quad 
\mathcal{F}_{IB} = 
\underbrace{I(X;Y)}_{\text{Bottleneck Size}} 
- \beta 
\underbrace{I(Y;U)}_{\text{Performance}}
\end{equation}
The constant $\beta$ sets the balance between bottleneck size and performance (Fig. \ref{fig:PIB_diagram}). If we set $\beta = 0$, the optimal point is wherever $I(X;Y) = 0$, but if we take $\beta \rightarrow \infty$, the optimal bottleneck size will be its maximum possible value, setting $X = Y$. In other words, shrinking the bottleneck is prioritized over performance when $\beta$ is small, and increasing performance is prioritized when $\beta$ is large. A stationary point of $\mathcal{F}_{IB}$ satisfies the following system of equations:
\begin{equation}\label{eqn:self_consistent}
\begin{aligned}
p(y|x) &= \frac{p(y)}{Z(\beta, x)}\exp \Big\{-\beta D_{kl}[P(U|x)|P(U|y)]\Big\}\\
p(y) &= \sum_{X}p(y|x)p(x) \\
p(u|y) &= \frac{1}{p(y)}\sum_{X}p(x,u)p(y|x)\\
\end{aligned}
\end{equation}
$D_{kl}[Q(X)|P(X)]$ is the KL-divergence, which is a difference measure between two distributions.
\begin{equation}
\displaystyle  D_{kl}[Q(X)|P(X)] \triangleq \sum_{X}q(x)\log\frac{q(x)}{p(x)} 
\end{equation}
$P(U|X)$ is a ``ground-truth" distribution, so the KL-divergence between $P(U|X)$ and $P(U|Y)$ can interpreted as a measure of $P(U|Y)$'s ``accuracy". $Z(\beta, x)$ is a normalizing constant for $P(Y|X)$. 

Stationary points are non-unique, nor is a global minimum guaranteed. Local solutions can be found by iteratively computing the minimum-search equations above, a method resembling the Blahut-Arimoto algorithm\cite{NIPS2003_gaussian_bottleneck}. Starting from initialized variables $p(y)$ and $p(u|y)$, $p(y|x)$ can be computed. This newly computed variable is then used to recompute the values of $p(y)$ and $p(u|y)$. Since we have new values of $p(y)$ and $p(u|y)$, we can repeat this cycle of computations until the values settle, arriving at a solution. One important contribution of this work is the transformation of these iterative minimum-search equations under the optimal closed-loop control context \eqref{eqn:continuous_iteration}\eqref{eqn:discrete_iteration}. Since $X$ and $Y$ accept many interpretations and encoding methods, the IB framework allows us to explore a wide range of algorithms that can generate a control signal.

\subsection{Path Integral (PI) Optimal Control}
Path integral control \cite{theodorouGeneralizedPathIntegral}\cite{williamsInformationTheoreticMPC2017} is a statistical optimal control formulation that incorporates more general problem formulations (e.g. nonlinear models and nonquadratic costs). The goal is to find a probability distribution over control signals $Q(U_{0:T})$ that minimizes a running cost. The optimization problem is as follows:
\begin{equation}\label{eqn:F_PI}
\begin{aligned}
% \min_{Q(U_{0:T})} \quad  \mathcal{F}_{PI} 
% &= \mathbb{E}_{Q}\Big\{\mathcal{S} -\lambda \log\frac{p(u_{0:T})}{q(u_{0:T})}\Big\}\\[0.1cm]
\min_{Q(U_{0:T})} \quad  \mathcal{F}_{PI} 
&= \underbrace{\mathbb{E}_{Q}\{\mathcal{S}\}}_{\text{State Penalty}} +\underbrace{\lambda D_{kl}[Q(U_{0:T})|P(U_{0:T})]}_{\text{Control Penalty}}\\[0.1cm]
\text{s.t.} \quad x_{t+1} &= f(x_t, u_t)\\
\mathcal{S} &= \sum_{t=0}^{T}\mathcal{L}(f(x_t, u_t)) = \sum_{t=0}^{T}\mathcal{L}(x_{t+1})
\end{aligned}
\end{equation}
where $x \in \mathbb{R}^{n_x},$ and $ u \in \mathbb{R}^{n_y}$. For joint variable time series, we use the shorthand $X_{0:T} \triangleq (X_0, X_1, ..., X_T)$. $\mathcal{S} \triangleq \mathcal{S}(u_{0:T}, x_0)$ is a running cost over the state trajectory. $\lambda > 0$ is a ``temperature" constant, adjusting how aggressive the optimization should be (Fig.\ref{fig:simple_entropy_example} and Equation (\ref{eqn:PI_solution})).  $P(U_{0:T})$ is a prior distribution. The intuition behind using $p(u_{0:T})$ in the control cost is discussed in the next section. 

As shown in \cite{williamsInformationTheoreticMPC2017}, the distribution $Q(U_{0:T})$ that minimizes $\mathcal{F}_{PI}$ can be expressed in closed-form.
\begin{equation}\label{eqn:PI_solution}
\begin{aligned}
 Q^{*}(U_{0:T}) &\triangleq \operatorname*{argmin}_{Q(U_{0:T-1})} \mathcal{F}_{PI}\\[0.1cm]
Q^{*}(U_{0:T} = u_{0:T}) &= q^*(u_{0:T}) \\
&= \psi^{*}p(u_{0:T})\\
\psi^{*} &\triangleq \psi^{*}(u_{0:T}, x_0) \triangleq \frac{1}{\eta}\exp\Big\{-\frac{\mathcal{S}}{\lambda}\Big\}
\end{aligned}
\end{equation}
where $\eta$ is a normalizing constant such that $\sum_{U_{0:T}}q^{*}(u_{0:T}) = 1$. $\psi^*$ can be thought of as ``cost weights" on $P(U_{0:T})$, determined by $\mathcal{S}$. If $\lambda \rightarrow 0$ in $\psi^{*}$, the distribution becomes more peaked, and as $\lambda \rightarrow \infty$, $\psi^{*}$ flattens out (Fig.\ref{fig:simple_entropy_example}). Note that $Q^{*}(U_{0:T})$ alone does not suggest a particular control law, since it is an open-loop optimization over distributions. In other words, the control realization $u_{0:T}$ is not a direct function of $\mathcal{S}$, $x_t$, or a set of parameters. We take advantage of this framing to open up a control-law-agnostic exploration\footnote{In practice, a control sequence is realized by taking an expected value over random samples, and re-optimizing every time step. While this computation could be considered a ``control law", it is not intrinsic to the path integral formulation. We do not take expected values to control a system, therefore maintaining a control-law-agnostic exploration.}.

\section{Path Integral Bottleneck}\label{sec:PIIB}
\begin{figure}
\includegraphics[scale=0.8, trim=0cm .5cm 0cm 0cm, center]{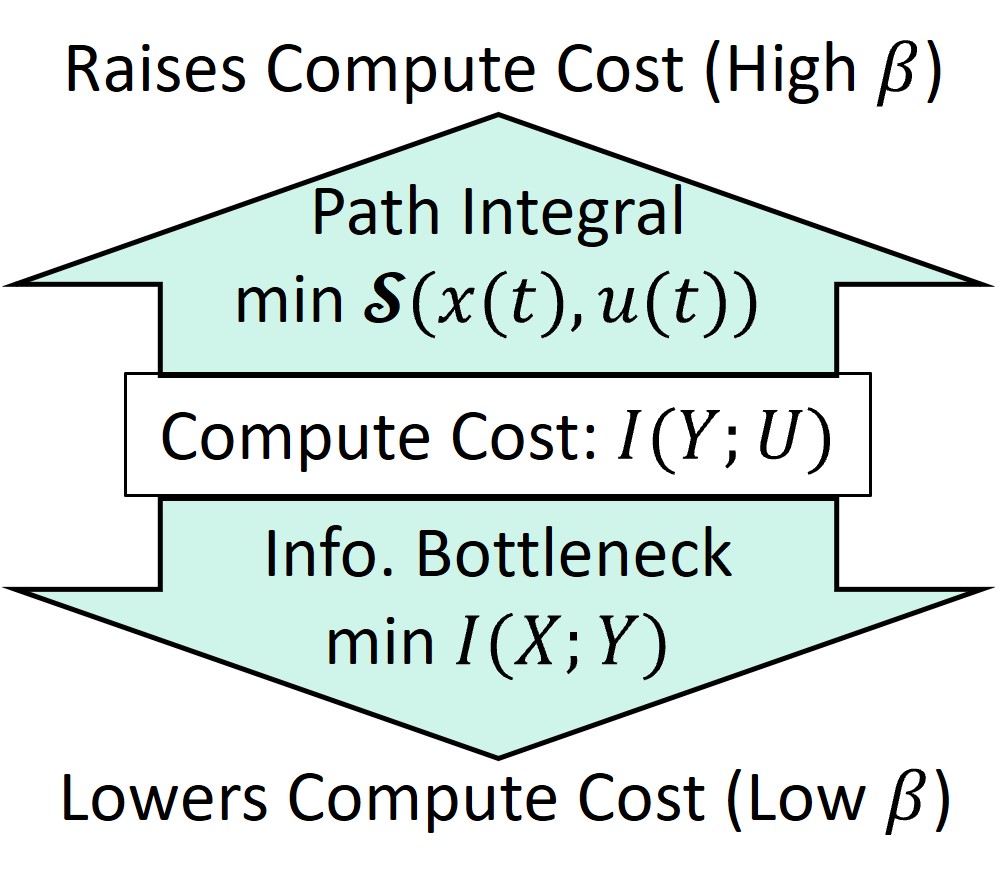}
\caption{The IB and PI affect compute cost in opposite directions. Optimizing the IB problem ($\min I(X;Y)$) lowers the compute cost, while optimizing the PI problem ($\min \mathcal{S}$) raises the compute cost. The parameter $\beta$ from the IB problem (\ref{eqn:F_IB}) determines which side is prioritized.}
\label{fig:PIB_diagram}
\vspace{-0.7cm}
\end{figure}
We present the PI-IB method, a combination of PI and IB that can quantify how sophisticated a feedback controller is by only observing behavior, excluding the controller's implementation details. PI uses the cost $\mathcal{S}(u_{0:T},x_0)$ to describe a controller's performance, but lacks a description of the its perception abilities. In other words, PI provides an offline statistical description of control. The IB provides perception context that is missing in the PI, but must do so with respect to the PI's running cost. We use this PI/IB synergy to quantify the Compute vs Control Cost tradeoff. However, the PI and IB describe open-loop systems, so they must be translated to a form that accurately describes closed loops. In this section, we first specify our definition of ``Compute Cost" for the open-loop, single-timestep case. We then close the loop to extend that definition to the multi-timestep case.

Our definition of compute cost is based on probability distributions over control signals $P(U)$. Without loss of generality, we will focus on distributions at $t = 0$, so $P(U_0)$. Since the PI and IB have different applications, they naturally admit multiple interpretations of $P(U_0)$. In the PI literature, $P(U_0)$ typically models literal actuator noise that is modeled with a Gaussian, $P(U_0) = \mathcal{N}(0, \Sigma_u)$. This noise distribution happens to be a good proxy for the control signal penalty ($u^\top \Sigma_u^{-1}u = u^\top Ru$). A low probability in the Gaussian distribution corresponds to a higher cost on $u$. By assuming that $P(U_0)$ is Gaussian, the LQR cost emerges from deriving the PI solution $q^{*}(u_0)$ (See Section \ref{sec:contiunous_variables} Equation (\ref{eqn:LQR_cost}).

However, the IB does not necessarily interpret the probability $P(U_0)$ as literal randomness. The mutual information $I(U_0;Y_0)$ in the IB problem is:
\begin{equation}
    I(U_0;Y_0) = H_P(U_0) - H(U_0|Y_0)
\end{equation}
% The interpretation that the prior is the "blind" behavior of the system is much less convincing than the interpretation that it's the state of knowledge of the system before compute. The computational irreducibility interpretation, hence, is stronger. Further emphasized with the hardware accelerator angle.
% The knowledge and perception ideas apply more to I(X;Y) than I(U;Y)
% \begin{equation}
% \begin{aligned}
% I(U_0;Y_0) &= \underbrace{H_P(U_0)}_{\text{Blind}}- \underbrace{H_P(U_0|Y_0)}_{\text{Perceiving}} & \text{IB interpretation}\\[0.2cm]
% &= \underbrace{H_P(U_0)}_{\text{Low Effort}} - \underbrace{H_Q(U_0)}_{\text{High Effort }} & \text{PI interpretation}
% \end{aligned}
% \end{equation}
Here, we can interpret entropy as lack of knowledge, and computation drives the system (and the system's potential observers) from a point of no knowledge, $H_P(U_0)$, to a point of increased knowledge, $H_P(U_0|Y_0)$. The point of no knowledge is prior to execution, when we cannot predict the control sequence. The point of increased knowledge is post-execution; the control sequence is realized and no more compute is needed. Since $I(U_0;Y_0)$ is the difference between these two points, it serves as a measure of compute effort. More specifically, it is the expected compute effort over $y_0$:
\begin{equation}\label{eqn:KL_IB}
I(U_0;Y_0) = \mathbb{E}_{P(Y_0)}\{D_{kl}[P(U_0|y_0)|P(U_0)]\}
\end{equation}
Each $y_0$ could be thought of as a separate task or objective for the controller. Since one task could be more complex than another, and therefore require different behavior, mutual information does not capture the compute effort of a specific task. We use $I(U_0|Y_0)$ for the average compute effort of a controller, and $D_{kl}[P(U_0|y_0)|P(U_0)]$ for the compute effort of a specific task. 
% This interpretation loosely follows the reasoning behind Kolmogorov randomness, which associates increased randomness with decreased compute. 
The mutual information $I(U_0;Y_0)$ can be more concretely interpreted as how coarsely-grained $U_0$ can be partitioned, or how much noise is removed from $U_0$. Both interpretations imply higher compute effort.

We can evaluate how effectively $P(U_0|Y_0)$ controls a system by setting $Q(U_0) = P(U_0|Y_0)$ in the PI problem (\ref{eqn:PI_solution}). This is intuitive, since $I(U_0|Y_0)$ should also represent the effort to lower the running cost $\mathcal{S}(u_0, x_0)$, which is enabled by sensory data $Y_0$. We can also set the ground-truth distribution to the optimal PI distribution, $P(U_0|X_0) = Q^{*}(U_0)$. A control decision that minimizes $\mathcal{S}$ should be a function of all the information available ($X_0$), whereas a control decision that uses partial information (e.g. $Y_0$ is a state estimate) does not necessarily minimize $\mathcal{S}$. 
\begin{equation}\label{eqn:prob_casting}
\begin{aligned}
p(u_0|x_0) &= q^{*}(u_0) = \psi^{*}p(u_0)\\
p(u_0|y_0) &= q(u_0) = \psi p(u_0)\\
\end{aligned}
\end{equation}
For the single timestep case, $\psi^* \propto \exp\{-\mathcal{L}(f(u_0, x_0))/\lambda\}$. Setting $P(U_0|Y_0) = Q(U_0)$ casts the Markov relationship $X_0 \rightarrow Y_0 \rightarrow U_0$ over the PI formulation. $P(U_0|X_0) = Q^{*}(U_0)$ casts a different dependency $Y_0 \rightarrow X_0 \rightarrow U_0$, which is necessary when solving the minimizing equations (\ref{eqn:self_consistent}). In the IB problem, the weights $\psi^{*}$ are used to compute the ground-truth $p(u_0|x_0)$, but $p(u_0|y_0)$ is used to compute $\psi$.

Although intuitive, these equalities are not intrinsic to the PI. Unlike the IB problem, PI is not concerned with how behavior changes with respect to state or representation. Since there is no notion of $P(X_0)$ or $P(Y_0)$ in the PI, $Q(U_0)$ is not a well-defined conditional probability. These distributions, along with $P(U_0|X_0)$ and $P(U_0|Y_0)$, are only well-defined in the IB. As a result, while the inequality $H_P(U_0|Y_0) \leq H_P(U_0)$ is guaranteed, $H_Q(U_0) \leq H_P(U_0)$ is not necessarily true. We must set $P(U_0)$ as a maximum entropy distribution to ensure $H_Q(U_0) \leq H_P(U_0)$. Overall, this interpretation casting between the PI and IB enables us to do a control-law agnostic analysis. We only look at the limits of how $Y_0$ influences the behavior $U_0$ without imposing a specific influence $Y_0$ has on $U_0$.

Defining the ground truth $P(U_0|X_0)$ is simple for a single-timestep cost. Since we are evaluating the complexity of an entire trajectory, we need the distribution $P(U_{0:T}|X_0) \triangleq P(U_0, U_1, ..., U_{T}|X_0)$. Using the PI solution (\ref{eqn:PI_solution}) and the PI to IB casting (\ref{eqn:prob_casting}), the ground-truth distribution\footnote{The influence of $P(U_{0:T})$ from (\ref{eqn:PI_solution}) is absent because we assume $P(U_{0:T})$ is uniform. Simply add a term in the exponent for the general case.} is the following, where $x_{t+1} = f(u_t, x_t)$:
\begin{equation}
p(u_{0:T-1}|x_0) \propto \exp\bigg\{-\sum_{t=0}^{T}\mathcal{L}( x_{t+1})/\lambda\bigg\}
\end{equation}\label{eqn:joint_def}
 The joint distribution can be decomposed by timestep using Bayes' Rule:
\begin{equation}\label{eqn:bayes_decomp}
\begin{aligned}
P(U_{0:T}|X_0) &= \prod_{t = 0}^{T} P(U_t|X_t) \\[0.2cm]
X_t &= (U_{0:t-1}, X_0) \hspace{0.4cm} \forall t > 0
\end{aligned}
\end{equation}

The distributions per timestep $P(U_t|X_t)$ can be expressed with respect to the Bayes' decomposition, yielding a value function expression.
\begin{equation}\label{eqn:bayes_decomp_two}
\begin{aligned}
p(u_t|x_t) \propto& \hspace{0.1cm} \exp\{-\mathcal{L}( x_{t+1})/\lambda\}W(u_t)  \\[0.2cm]
W(u_t) \triangleq& \sum_{U_{t+1:T}} \exp\bigg\{-\sum_{\tau=t+1}^{T}\mathcal{L}(x_{\tau+1})/\lambda\bigg\}
\end{aligned}
\end{equation}
Being able to define each individual $P(U_t|X_t)$ means we can formulate an IB cost for each timestep.
\begin{equation}\label{eqn:timestep_F_IB}
\begin{aligned}
\min_{\substack{P(Y_t|X_t),\\ P(Y_t), \\ P(U_t|Y_t)}} \quad \mathcal{F}_{IB}^{(t)} &\triangleq I(X_t;Y_t) - \beta I(U_t;Y_t)
\end{aligned}
\end{equation}
The PI-IB method applies minimum search to (\ref{eqn:timestep_F_IB}) for various $\beta$ and $t$ to quantify the compute-control tradeoff of an online closed-loop controller with some ability to plan, as opposed to an offline IB $\mathcal{F}_{IB} = I(X_0;Y_0) - \beta I(U_{0:T-1};Y_0)$ that suggests the entire trajectory is planned in one timestep. Separating the joint IB by timestep also circumvents the curse of dimensionality when computing the minimum-search equations. Using this closed-loop IB, we derive the single-timestep equations needed to find a minimum of the IB cost (\ref{eqn:timestep_F_IB}) in the following two sections, one for the continuous-variable (Section \ref{sec:contiunous_variables}), LQR case and one for the discrete-variable, nonlinear case (Section \ref{sec:discrete_variables}).

% Explain intuitively how the cost function iteration is the key to all this (for example, a flat cost function easily corresponds to no compute), Set $\beta$ to 0, set $\lambda$ to infinity, what happens? What does it say about $P(U)$ and $Q(U)$

% PIB is being run at every timestep, a one-step open loop optimization is basically a closed-loop control strategy. Again, this can be extended in future works with the same formulation. 

% Communication and noise as well.
% $P(Y|X)$ could just be noise, but it could also be other representations. 
% Note that communication, compute, and representation are all the same thing here. So here we interpret it as compute. 

% \begin{equation}
% H(U_\psi) = -\sum_U e^{-\frac{S(u)}{\lambda}}p(u) \log e^{-\frac{S(u)}{\lambda}}p(u)
% \end{equation}

% Note that $\Psi(U=u) = \Psi(u)$ is an abuse of notation. $\Psi$ is not a probability distribution, as $\Psi$ must be normalized with respect to $Q$.

\vspace{-0.1cm}
\section{Continuous Variables}\label{sec:contiunous_variables}
In this section, we derive the minimum-search IB equations (\ref{eqn:continuous_iteration})
and a way to recover the resulting control distribution (\ref{eqn:sigma_u|y}) for an LQR system, allowing use to evaluate the compute-control tradeoff. We restrict our maximum-entropy prior distributions ($P(X_t)$, $P(U_t)$) to Gaussians. Over the set of continuous probability density functions, a maximum entropy distribution cannot be well-defined without moment constraints (e.g. covariance), and Gaussians are easily defined by this constraint.

Previous work \cite{NIPS2003_gaussian_bottleneck} shows that when the prior distributions are Gaussian ($P(X) = \mathcal{N}(0, \Sigma_{x})$, $P(U) = \mathcal{N}(0, \Sigma_u)$, assume zero mean without loss of generality), the optimal intermediate variable $y$ in the Markov relationship $X\rightarrow Y\rightarrow U$ is a linear transform of $x$.  
\begin{equation}\label{eqn:linear_observer}
    y = Cx + \xi \hspace{1cm} \xi \sim\mathcal{N}(0, \Sigma_\xi)
\end{equation} 
$x$ and $y$ are the same dimension, so $x, y \in\mathbb{R}^{n_x}$ and $u \in\mathbb{R}^{n_u}$. All covariance matrices are positive definite. Since we know $y$ takes this form at the optimal point of the IB function $\mathcal{F}_{IB}$, we can optimize over the parameters $C$ and $\Sigma_{\xi}$ instead of directly iterating over all possible distributions $P(Y|X)$.
\begin{equation}
\begin{aligned}
\min_{C,\Sigma_\xi} \quad & \mathcal{F}_{IB} = I(X;Y) - \beta I(Y;U) \\
\text{s.t.} \quad & y = Cx + \xi  \\
& \xi\sim\mathcal{N}(0, \Sigma_\xi)\\
\end{aligned}
\end{equation}
To fulfill the stationary point conditions (\ref{eqn:self_consistent}), we can perform the following iterations as described in \cite{NIPS2003_gaussian_bottleneck}, where the iteration index is $k$.
\begin{equation}\label{eqn:continuous_iteration}
\begin{aligned}
\Sigma_{y_k} &= C_{k}\Sigma_{x}C_{k}^\top + \Sigma_{\xi_k}\\
% \Gamma_k &= \Sigma_{uy_k}\Sigma_{y_k}^{-1}\\
% \Sigma_{y_k|u}^{-1}&= \Sigma_{y_k}^{-1} + \Gamma_k^\top\Sigma_{u|y_k}^{-1}\Gamma_k\\
\Sigma_{y_k|u}&= C_k\Sigma_{x|u}C_k^\top + \Sigma_{\xi_{k}}\\
\Sigma_{\xi_{k+1}} &= (\beta\Sigma_{y_k|u}^{-1}-(\beta-1)\Sigma_{y_k}^{-1})^{-1}\\
C_{k+1} &= \beta\Sigma_{\xi_{k+1}}\Sigma_{y_k|u}^{-1}C_{k}(I-\Sigma_{x|u}\Sigma_{x}^{-1})\\
\end{aligned}
\end{equation}
where $I$ is the identity matrix. In order to perform these iterations, the conditional covariance matrix $\Sigma_{x|u}$ needs to be defined, and in general the conditional distribution between $U_t$ and $X_t$ needs to be defined as a prior. We define $\Sigma_{x|u}$ with the discrete-time linear model and quadratic cost:
\begin{equation}\label{eqn:LQR_setup}
\begin{aligned}
x_{t+1} &= Ax_t + Bu_t\\
\mathcal{L}(x) &= \frac{1}{2}x^\top Qx
\end{aligned}
\end{equation}
$Q$ is symmetric positive semidefinite. If we expand the PI solution (\ref{eqn:PI_solution}) for one timestep, we get the following. 
\begin{equation}\label{eqn:LQR_cost}
\begin{aligned}
q^{*}(u_t) &\propto \exp \Big\{-\frac{1}{2\lambda}x^\top_{t+1}Q x_{t+1}\Big\}p(u_t)\\
&= \exp\Big\{-\frac{1}{2}(\frac{1}{\lambda}x^\top_{t+1}Q x_{t+1} + u^\top_t\Sigma_u^{-1}u_t)\Big\}\\
\end{aligned}
\end{equation}
When accounting for a longer horizon, which means including the variable $W(u_t)$ in (\ref{eqn:bayes_decomp_two}), replace $A$, $B$, and $Q$ with augmented matrices for the remaining derivation. The quadratic state cost naturally combines with the Gaussian prior to produce the LQR cost where $R=\Sigma_{u}^{-1}$, penalizing high values of $u$\cite{theodorouGeneralizedPathIntegral}. Furthermore, the Gaussian priors also produce a linear observer (\ref{eqn:linear_observer}) without having to explicitly define it. Next, we use the linear model (\ref{eqn:LQR_setup}) to expand the expression. We set $\widetilde{Q} \triangleq \frac{1}{\lambda}Q$.
\begin{equation}\label{eqn:expanded_conditional}
\begin{aligned}
q^{*}(u_t) &\propto \exp\Big\{-\frac{1}{2}(x^\top_{t+1}\widetilde{Q} x_{t+1} + u^\top_t\Sigma_u^{-1}u_t)\Big\}\\
&= \exp\Big\{-\frac{1}{2}(u^\top_t(\Sigma_u^{-1}+ B^\top \widetilde{Q}B)u_t \\
&\hspace{2cm}+ 2x_t^\top A^\top \widetilde{Q}Bu_t \\
&\hspace{2cm}+ x_t^\top A^\top \widetilde{Q} Ax_t)\Big\}\\
\end{aligned}
\end{equation}
Since we are casting the interpretation $P(U_t|X_t)=Q^{*}(U_t)$ over the IB, we can get a joint distribution using Bayes Rule. By taking the expansion (\ref{eqn:expanded_conditional}) and multiplying it with $P(X_t)$, we get the joint distribution:
\begin{equation}\label{eqn:PI_covariance}
\begin{aligned}
 P(U_t,X_t) &= \mathcal{N}\Bigg(\begin{bmatrix} A^\top \widetilde{Q}A + \Sigma_{x}^{-1} \hspace{-10pt}& A^\top \widetilde{Q}B\\ B^\top \widetilde{Q}A \hspace{-10pt} & B^\top \widetilde{Q}B + \Sigma_{u}^{-1}\end{bmatrix}^{-1}\Bigg)
\end{aligned}
\end{equation}
In terms of covariance matrices, the joint distribution is:
\begin{equation}\label{eqn:main_joint_covariance}
\begin{aligned}
P(U_t, X_t) &= \mathcal{N}\Bigg(\begin{bmatrix} \Sigma_{x} & \Sigma_{xu} \\ \Sigma_{ux} & \Sigma_{u}\end{bmatrix}\Bigg)\\
\end{aligned}
\end{equation}
where $\Sigma_{xu}$ and $\Sigma_{ux}$ are the corresponding cross-covariance matrices. The means are zero-vectors. The Schur complement of $\Sigma_{u}$ is $\Sigma_{x|u}$. By expressing the inverse of the joint covariance in (\ref{eqn:main_joint_covariance}) with the Schur complement and equating it to the matrix derived in  (\ref{eqn:PI_covariance}), we get the following equality, giving us the conditional needed for the minimum-search equations. 
\begin{equation}
\Sigma_{x|u} = (A^\top \widetilde{Q}A + \Sigma_x^{-1})^{-1}
\end{equation}
Using the same process, we can also get the equality $\Sigma_{u|x} = (B^\top \widetilde{Q}B + \Sigma_u^{-1})^{-1}$. Since we do not define $P(U_t|X_t)$ as an explicit conditional probability, but rather an interpretation following the PI method, it is worth checking whether $H_P(U_t) \geq H_P(U_t|X_t) = H_{Q^{*}}(U_t)$ so that a fundamental law of entropy is not violated. For Gaussian distributions, the differential entropies are related such that:
\begin{equation}
\begin{aligned}
H_P(U_t) &\propto \log(\det(\Sigma_{u}))\\
H_P(U_t|X_t) &\propto \log(\det(\Sigma_{u|x}))\\
&= \log(\det((B^\top \widetilde{Q}B + \Sigma_u^{-1})^{-1}))
\end{aligned}
\end{equation}
In order for $H_P(U_t) \geq H_P(U_t|X_t)$, the following must hold:
\begin{equation}
\det(\Sigma_{u}) \geq \det((B^\top \widetilde{Q}B + \Sigma_u^{-1})^{-1})
\end{equation}
Invert both sides to reflect the inequality:
\begin{equation}
\det(B^\top \widetilde{Q}B + \Sigma_u^{-1}) \geq \det(\Sigma_{u}^{-1})
\end{equation}
Since $\Sigma_u^{-1}$ is positive definite and $B^\top \widetilde{Q}B$ is positive semi-definite, this inequality holds (Minkowski determinant inequality), preserving the conditional entropy inequality.

While $\Sigma_{u|y}$ is not necessary for the minimum-search equations, we must recover it to compute $\mathcal{F_{PI}}$ for the continuous variable case. From the Schur inverse of the joint covariance matrix between $u$ and $y$ we get the following:
\begin{equation}\label{eqn:sigma_u|y}
\Sigma_{u|y} = (\Sigma_u^{-1} + \Sigma_u^{-1}\Sigma_{uy}\Sigma_{y|u}^{-1}\Sigma_{yu}\Sigma_u^{-1})^{-1}
\end{equation}
$\Sigma_{uy}$, $\Sigma_{yu}$, $\Sigma_{ux}$, and $\Sigma_{xu}$ can be found by using (\ref{eqn:linear_observer}), (\ref{eqn:PI_covariance}), and (\ref{eqn:main_joint_covariance}).
\begin{equation}
\begin{aligned}
\Sigma_{uy} &= \Sigma_{ux}C^\top & \Sigma_{yu} &= C\Sigma_{xu}\\
\Sigma_{ux} &= -\Sigma_{u}B^\top A^{-\top} & \Sigma_{xu} &= -A^{-1}B\Sigma_{u}\\
% \Sigma_{ux} &= -\Sigma_{u}B^\top A\Sigma_{x|u} &\Sigma_{xu} &= -\Sigma_{x|u}A^\top B\Sigma_u\\
% \Sigma_{u|x}& = (A^\top \widetilde{Q}A+\Sigma_x^{-1})^{-1} &\Sigma_{x|u} &= (B^\top \widetilde{Q}B+\Sigma_u^{-1})^{-1} 
\end{aligned}
\end{equation}
When setting $\Sigma_{\xi} = 0$ and setting $C = I$, we see $\Sigma_{u|y} = \Sigma_{u|x}$. If $C = 0$, then $\Sigma_{u|y} = \Sigma_{u}$. So when $C$ is interpolated between no observation to full state observation, the control distribution interpolates between the maximum entropy and ground-truth distributions.

\vspace{-0.5cm}
\section{Discrete Variables}\label{sec:discrete_variables}
\vspace{-0.1cm}
We now consider the discrete case, which allows more flexibility in our model and cost definitions. The maximum entropy point is also much easier to define. For any distribution of support size $N$, the maximum entropy distribution is the uniform distribution. Therefore, the following distributions can be defined.
\begin{equation}\label{eqn:uniform_definitions}
\begin{aligned}
\text{Priors:} \quad & p(x_t) = \frac{1}{N} \hspace{1cm} p(u_t) = \frac{1}{M} \\[0.2cm]
\text{Ground Truth:} \quad & p(u_t|x_t) = \frac{\psi_t^{*}}{M} \hspace{1cm} p(u_t,x_t) = \frac{\psi_t^{*}}{MN}
\end{aligned}
\end{equation}
% \begin{equation}\label{eqn:uniform_definitions}
% \begin{aligned}
% p(x) &= \frac{1}{N} \hspace{1cm} p(u) = \frac{1}{M}\\[0.2cm]
% p(u|x) &= \frac{\psi_{xu}^{*}}{M} \hspace{1cm} p(u,x) = \frac{\psi_{xu}^{*}}{MN}\\
% \end{aligned}
% \end{equation}
$N$ and $M$ are the support sizes for the random variables $X_t$ and $U_t$. $\psi^{*}_t \triangleq \psi^{*}(u_t, x_t)$ has the same purpose $\psi^{*}$ in the PI solution (\ref{eqn:PI_solution}), acting as the set of ``cost weights". However, the factor $W(u_t)$ is included from the Bayes' decomposition (\ref{eqn:bayes_decomp_two}), hence the subscript $t$. The shape of $\psi^{*}_t$ can be arbitrary as long as $\sum_{U}p(u|x) = 1$. The IB minimum-search algorithm (\ref{eqn:self_consistent}) simplifies to the following with uniform prior distributions. The $t$ subscript is dropped for $x_t, y_t,$ and $u_t$.
\begin{equation}\label{eqn:discrete_iteration}
\begin{aligned}
p(y_{k+1}|x) &\displaystyle = \frac{p(y_k)}{Z(\beta, x)}\exp\Big\{\frac{\beta}{M}\sum_{U}\psi^*_t \log\frac{\psi(y_k,u)}{\psi^*_t }\Big\}\\[0.1cm]
p(y_{k+1}) &= \frac{1}{N}\sum_{X}p(y_{k+1}|x)\\[0.1cm]
p(u|y_{k+1}) &= \frac{\psi(y_{k+1},u)}{M}\\[0.2cm]
\psi(y_k,u) &\triangleq \sum_{U}p'(y_k|x)\psi^*_t\\[0.1cm]
p'(y_k|x) &\triangleq\displaystyle \frac{p(y_k|x)}{\sum_X p(y_k|x)}
\end{aligned}
\end{equation}
The weights $\psi(y_k, u)$ correspond to a proxy value function $\psi(y_k,u) = \eta^{-1}\exp\{-\lambda^{-1}\mathcal{L}_k(f'(y_k,u))\}W_k(u_t)$ that is modified every iteration, following the definition in (\ref{eqn:PI_solution}). Computing $\psi(y_k,u)$ is equivalent to a matrix multiplication on $\psi_t^{*}$. If $\psi(y,u) = 1$, then the bottleneck can be collapsed to 0 since all information from $\psi_t^{*}$ is destroyed, but if $\psi(y,u) =\psi_{t}^{*}$ then $Y$ would have to represent $X$ more exactly, widening the bottleneck. Thus, finding an optimal bottleneck size corresponds to finding an ideal compromise between $\psi(y,u) = 1$ and $\psi(y,u) =\psi_t^{*}$.

\section{Simulation}\label{sec:simmy}
We apply PI-IB to evaluate the compute-control tradeoffs for the cart-pole problem in MATLAB (Fig.~\ref{fig:cost_percent}). The cart-pole problem provides a wide range of behaviors to analyze. Simple balancing, which PID, LQR, or Bang-Bang can accomplish, does not require planning. A swing-up task, which MPC accomplishes, does require planning. PI-IB allows us to put these behaviors on a continuum, revealing the different compute demands of each task agnostic to algorithm details. 

\subsection{Setup}
The running cart-pole cost we use is $\mathcal{L}(x_{t}) = 500\phi_{t}^2 + \dot{\phi}_{t}^2 + 10\dot{q}_{t}^2 + 5q_{t}^2$, where $x = [\phi\hspace{0.2cm} \dot{\phi} \hspace{0.2cm} q \hspace{0.2cm} \dot{q}]^\top$ is the state vector, $\phi$ is the angle away from the balance point (up position is $0^{\circ}$), and $q$ is the cart position. We use the discrete variable formulation, since the uniform prior distribution easily encodes that we expect the cart-pole to succeed for $N$ initial states. We set the temperature variable to $\lambda = 0.2$, usually set with some arbitrariness.

\begin{figure}
\includegraphics[scale=0.62, trim=3.5cm 9cm 0cm 8cm]{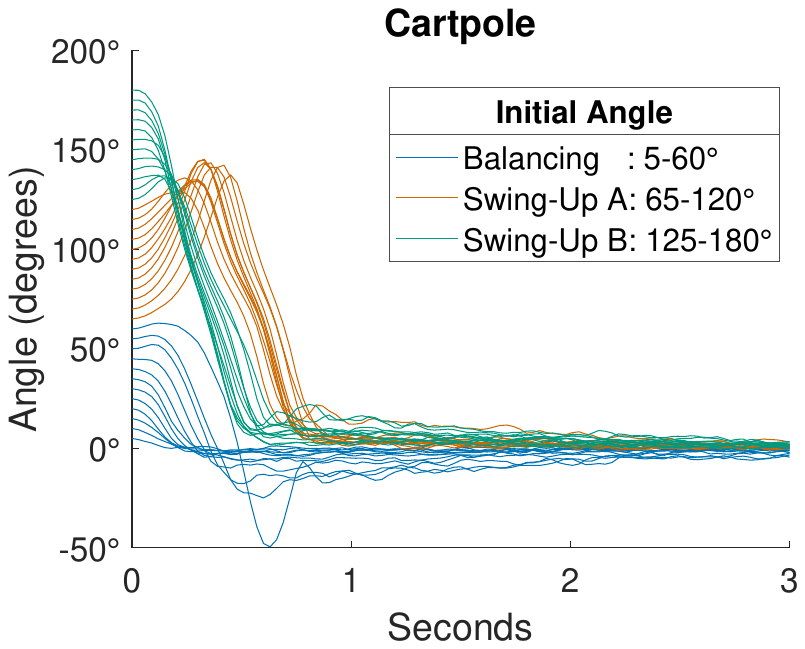}
\caption{Results of cartpole control with MPPI, with starting angles 5-180$^{\circ}$. $0^{\circ}$ is the balance point. All controls successfully drive the angle to the balance point. Transitioning from the balancing task to the swing-up task is visually distinct. We name the three distinct regions ``Balancing" (5-60$^{\circ}$), ``Swing-up A" (65-120$^{\circ}$), ``Swing-up B" (125-180$^{\circ}$). The discrete timestep for the control is .03s.}
\label{fig:cartpole_trajectories}
\vspace{-0.1cm}
\end{figure}

\begin{figure}
\includegraphics[scale=0.44, trim=0.9cm 8.5cm 0cm 8cm]{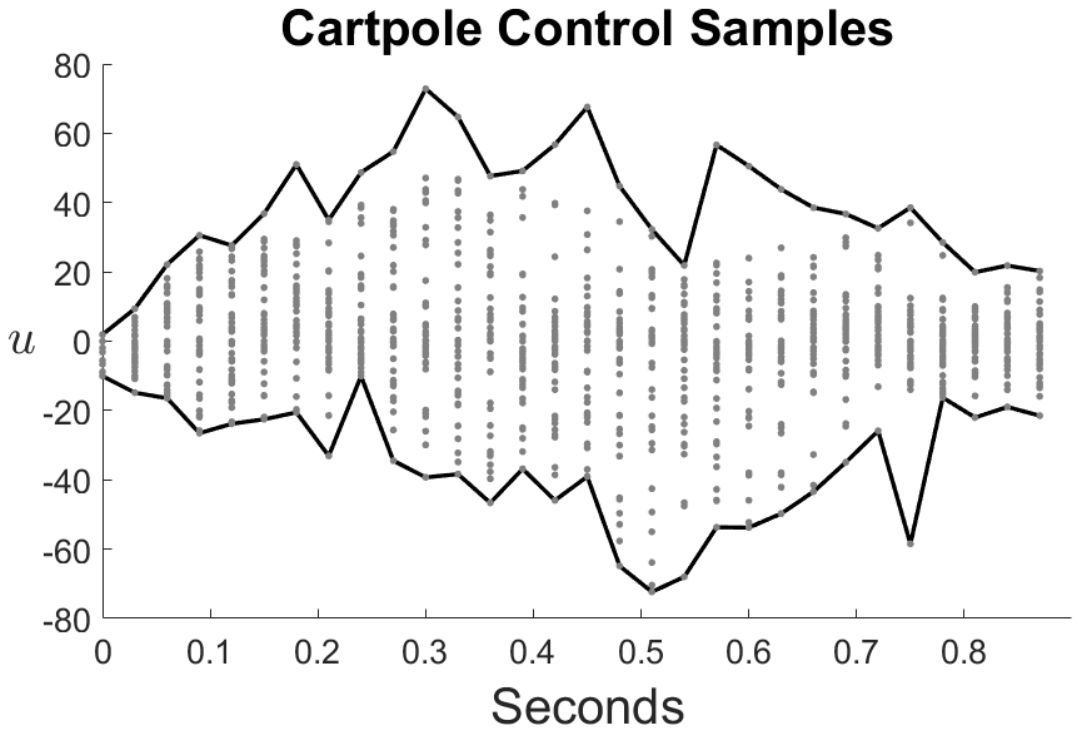}
\caption{Random samples (7188 shown) of viable cart-pole control across initial angles $5-180^{\circ}$. Points are sparser towards the envelope and denser towards towards the center. Some timesteps have a narrower range than others. This emergent shape is evidence of a sophisticated controller.}
\label{fig:cartpole_samples_visual}
\vspace{-0.5cm}
\end{figure}

In order to run the IB equations (\ref{eqn:discrete_iteration}), we must construct the ground-truth distributions $P(U_t|X_t)$. Fully searching the space of trajectories is impractical, so we must have a search strategy to build a set of viable trajectories. We first run MPPI \cite{williamsInformationTheoreticMPC2017} on the cartpole with initial angles ranging from 5-180$^{\circ}$ (Fig.\ref{fig:cartpole_trajectories}). Then, we take the resulting control signals and add Gaussian variations. Variations are pruned if that variation increases the cost too much. Otherwise, it is added to the set of viable variations. Fig. \ref{fig:cartpole_samples_visual} shows the resulting shape of the control signals after the samples are collected. While it is clear that the task can afford randomness, its distinct shape is evidence that sophisticated control is present.
We find the minimum and maximum values over all control signals in the set, which defines the upper and lower bound of the random variable $U_t$. The signals are discretized by increments of $0.3$, since smaller increments do not affect the results. All prior distributions $P(U_t)$ are a uniform distribution over these increments. By running this set of controls over all initial angles, the ground-truth $P(U_t|X_t)$ can be constructed based on the discretization. We take a horizon of 0.9 seconds, where all the trajectories get close to balance point, and decompose it into $T = 30$ timesteps, which gives us 30 IB problems per the Bayes' decomposition (\ref{eqn:bayes_decomp}, \ref{eqn:bayes_decomp_two}). $P(U_t|X_t)$ is computed in a receding horizon fashion, depending on states $t:T$. We set $p(x_t) = 1/N$ for all the IB problems, where $N$ is the number of initial conditions. This does not need to change with time, since at each timestep the state is already determined before the next control signal is generated. On the first IB iteration $k=1$, we set $p(y_t) = 1/N$ and $p(u_t|y_t) = p(u_t|x_t)$.

\subsection{Results}
% And in the results, start with the results that most support your statements in the intro (look, we have characterized compute cost and it is as we expect: Fig 6 and 8), and end with results that are less intuitive (e.g. Fig7)
\begin{figure}
\includegraphics[scale=0.51, trim=2.1cm 16cm 0cm 2cm]{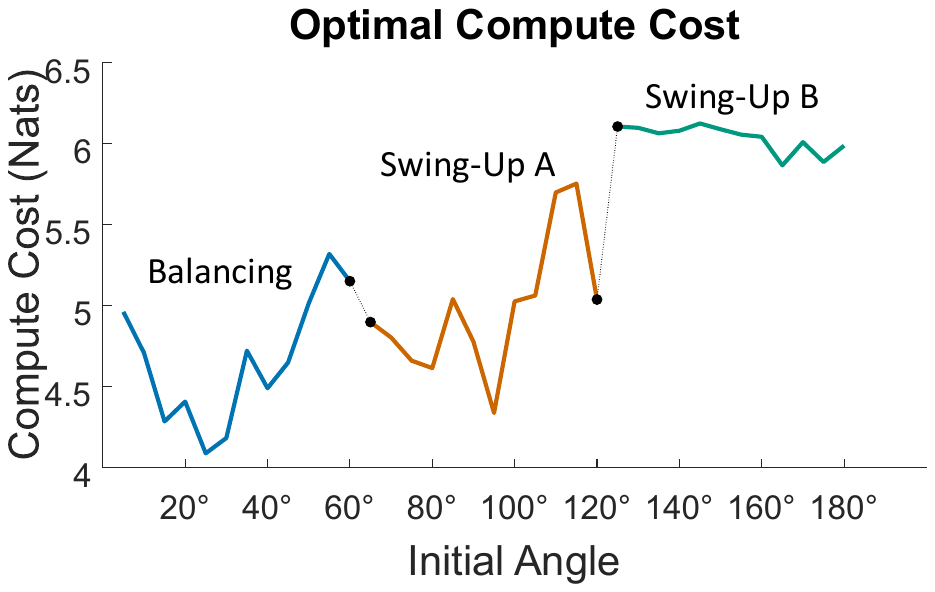}
\caption{Compute cost with the best performance (minimum cartpole cost) for each initial angle. Compute cost at the larger initial angles are significantly higher, indicating less room for randomness. Colors correspond to ``Balancing", ``Swing-Up A", and ``Swing-Up B" behavior.}
\label{fig:optimal_compute_cost}
\vspace{-0.3cm}
\end{figure}

At the IB cost's stationary point, $\beta$ prioritizes task performance over bottleneck size (Fig. \ref{fig:PIB_diagram}, equation (\ref{eqn:F_IB})). For each IB problem we sweep $\beta \in [0.9, 1]$. The IB tradeoff converges when $\beta < 0.9$ or $\beta > 1$, so going outside this range is unnecessary. The compute cost per initial angle is found by computing the resulting KL-divergence (\ref{eqn:KL_IB}) per angle. Since we have a different IB problem per timestep, we take the average compute cost over time. The expected control cost per compute cost is evaluated by feeding the expected control signal into the running cost: $\sum_{t=0}^{T}\mathbb{E}_{P(U_t|y_t)}\{\mathcal{L}(x_{t+1})\}$.

In Fig. \ref{fig:optimal_compute_cost}, the compute cost that minimizes the control cost is plotted against initial cart-pole angle. While there isn't a monotonic trend between small angle increments, we see a broader trend when we partition the space into three ranges $55^{\circ}$ each. This partitioning matches the noticeable behavior differences that emerge in the cart-pole trajectories (Fig.\ref{fig:cartpole_trajectories}), which we name ``Balancing" (5-60$^{\circ}$), ``Swing-up A" (65-120$^{\circ}$), and ``Swing-up B" (125-180$^{\circ}$). The range of compute costs for ``Balancing" overlaps with ``Swing-Up A", with ``Swing-Up A" having higher compute cost. ``Swing-Up B" has a noticeably larger compute cost. Intuitively, the tasks at high initial angles are more ``difficult" than the tasks that start off close to the balance point, since starting further away from the balance point requires more planning.

\begin{figure}
\includegraphics[scale=0.45, trim=1cm 9.8cm 0cm 2cm]{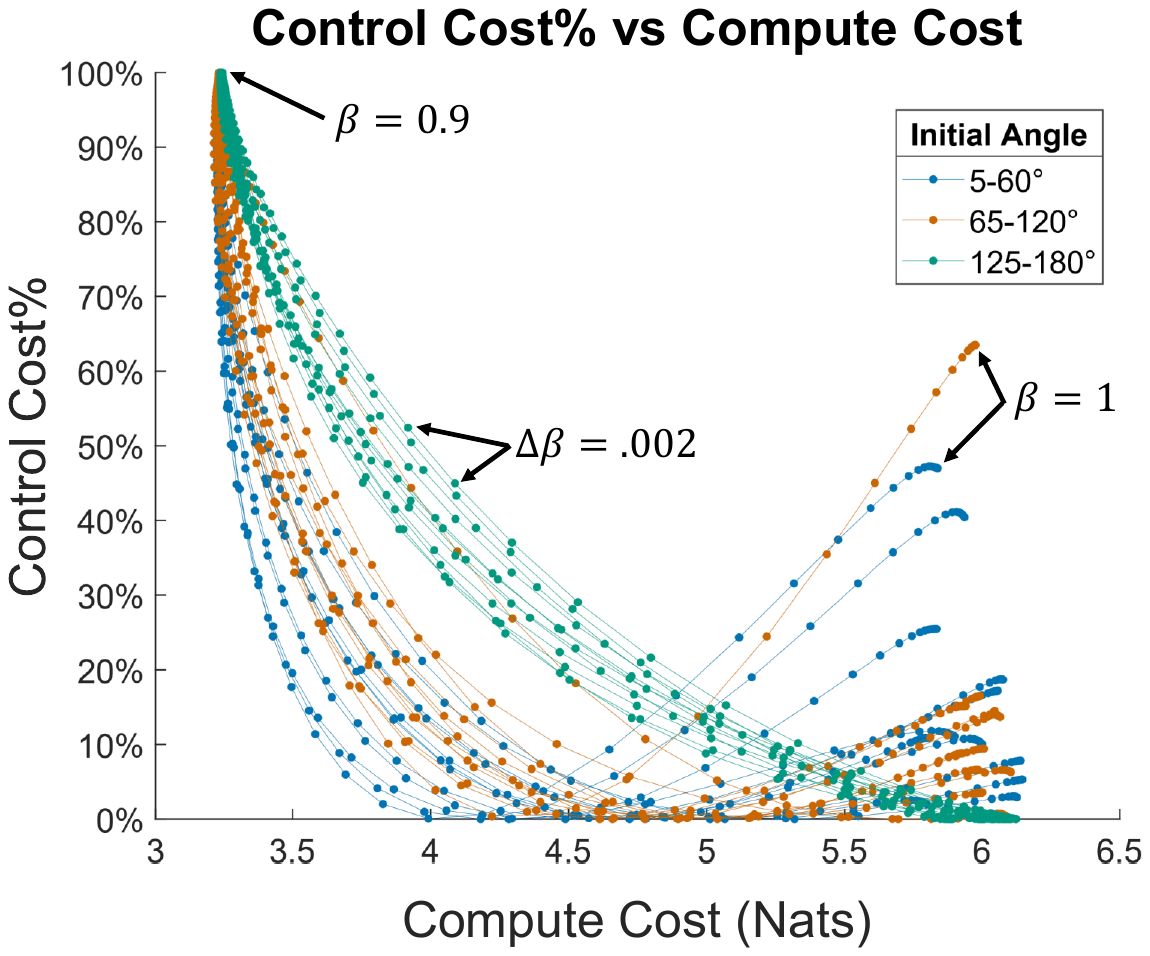}
\caption{Control vs Compute cost curves for initial cartpole angles 5-180$^{\circ}$, step size 5$^{\circ}$. Markers represent values of $\beta$, swept from 0.9 to 1, step size .002. Cost is normalized to a percentage, where $100\%$ is the maximum expected cost computed in the sweep, and $0\%$ is the minimum.}
\label{fig:cost_percent}
\vspace{-0.7cm}
\end{figure}

Fig. \ref{fig:cost_percent} draws the compute-control tradeoff over the sweep $\beta \in [0.9, 1]$ for each initial angle. The range of the costs over different initial angles is too large to plot properly, so the costs are normalized between their minimum (0\%) and maximum (100\%) values. Despite each point in Fig.~\ref{fig:cost_percent} corresponding to evenly spaced increments of $\beta$, the points are sparsest in the middle while densest at the two endpoints. Furthermore, the slopes between each point are steep when compute effort is low. This trend indicates significant improvements can be made on the control task in the region where compute effort is low. However, marginal improvements decrease as compute effort increases, suggesting that after a certain point additional compute effort is wasteful. 

Once again, the curves are noticeably different between the three behaviors. ``Balancing" and ``Swing-Up A" have overlapping curves, but in general ``Balancing" has steeper slopes at the low compute cost region. A more distinct difference is that ``Balancing" and ``Swing-Up A" have many curves that after descending to a minimum point, the curves trend upwards, whereas all curves in ``Swing-Up B" monotonically decrease. At first glance, this is contrary to the intuition we built early, where control cost should decrease with compute cost. We attribute this to a form of overfitting. In other words, randomness is beneficial in these regions, while ``Swing-Up B" is the group that affords the least randomness.

% The curves in Fig.~\ref{fig:cost_percent} provide a fundamental characterization of models and control cost functions, allowing us to predict and compare how certain control sequences and strategies would perform. 
% It's percent, NOT cost, think about the RANGE of the compute cost rather than the raw values.
% Roughly shift the gap
% For example, Bang-Bang control would have $H(U) = \log(2) \approx 0.7$, which serves as an upper bound on $I(U;Y)$. We can estimate that there is some granularity where it does ok.
% Cost swings too much per initial condition, hard to say.

\begin{figure}
\includegraphics[scale=0.49, trim=2.5cm 6.8cm 0cm 5cm]{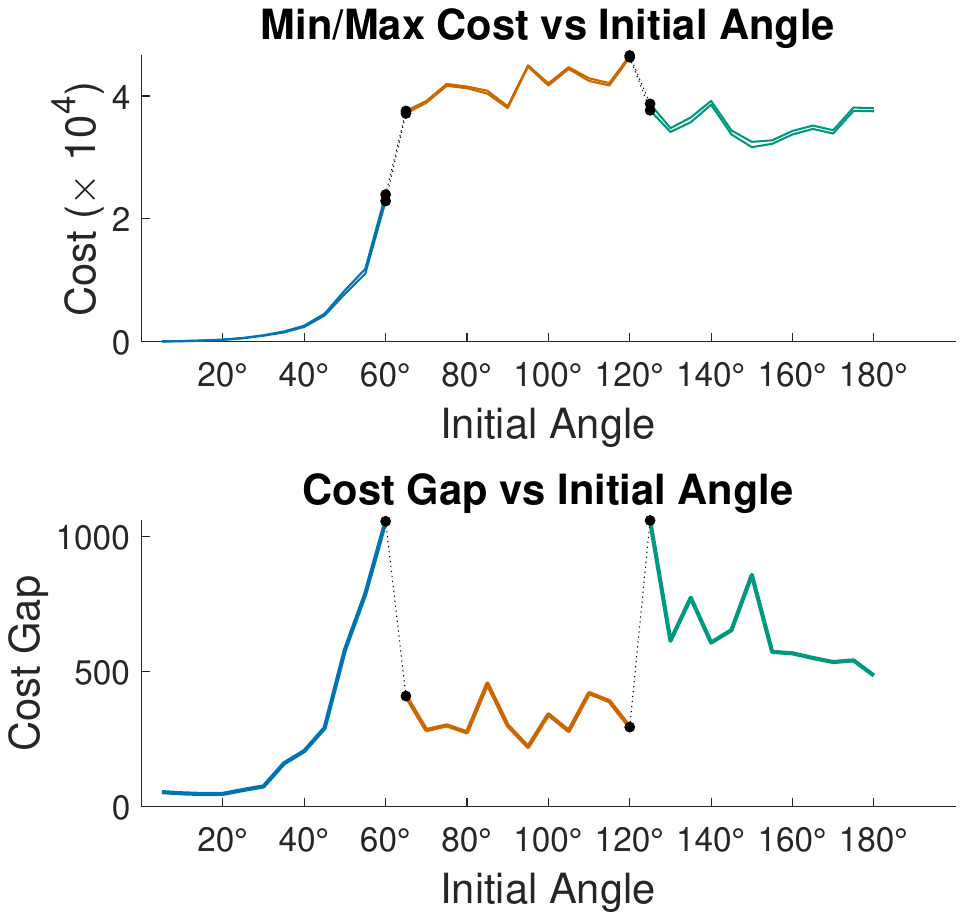}
\caption{Scale of expected control costs over the cartpole's trajectory vs initial cartpole angles. The expected cost over $P(U|Y)$ starts off at the maximum when $\beta = 0.9$. The minimum cost is reached within $\beta \in [0.9, 1]$. The ``Cost Gap" is the min-max difference.}
\label{fig:max_costs}
\vspace{-0.7cm}
\end{figure}
Fig.\ref{fig:max_costs} shows the scales of the control costs when sweeping $\beta \in [0.9, 1]$. The cost and cost gap increase monotonically in the ``Balancing" region, as opposed to ``Swing-Up A/B". When plotting the cost gap, which is the maximum decrease in control cost when sweeping $\beta \in [0.9, 1]$, ``Swing-Up A" is noticeably small. From Fig.\ref{fig:cartpole_trajectories}, the ``Swing-Up A" behavior is the last to arrive at the balance point, which may put a lower bound on the cost and limit the cost gap.

\section{Conclusions and Future Work}
In this work, we combine the PI and IB methods (PI-IB) to explore the relationship between task performance and compute cost.
%, using the interpretation of entropy as lack of knowledge before an algorithm is run. 
Our approach allows for a flexible analysis that is not bound to a particular encoding, due to the IB, and not bound to any particular control law or algorithm, due to the PI. Simulations of a cart-pole support the intuition that higher computation capacity can decrease the control cost, and that more difficult tasks require more compute effort. Future investigation could explore the conditions under which complexity increases or decrease with time. This flexibility will allow us to analyze the behavior of a biological control system for cross-platform comparisons.

% In this work we find that the compute-control tradeoff varies with time, and while we simply take the average over time, there is still room to investigate what the time-dependent information bottleneck can say about the system. A varying tradeoff over time suggests that the system moves between high-entropy and low-entropy regions of the state space.

% In this work we made maximum entropy assumptions on prior distributions, but through data analysis, adding constraints, or adding more system knowledge, priors can be modified. The more low entropy a prior is, the more specialized the system. Thus, there is potential for the PIB framework to compare efficiency between specialized and general algorithms or computing systems. Furthermore, PIB could provide a tool for analyzing data and determining bounds for alternative, embodied computing systems, as well as interpret computation in other biological systems.

% MPC analysis, bottleneck for two types of horizons…? Analysis over different entropy regions (startup vs unstable equilibrium). Linear System, find global minimum. Find connection between uniform distribution and Linfinity control

% Notable is that both the PI and IB are free-energy minimization problems (see Equation (30) in \cite{tishby2000informationbottleneckmethod} and Equation (12) in \cite{williamsInformationTheoreticMPC2017}). 

% \input{sections/temporary}

\bibliography{citations}

\begin{thebibliography}{10}
\providecommand{\url}[1]{#1}
\csname url@samestyle\endcsname
\providecommand{\newblock}{\relax}
\providecommand{\bibinfo}[2]{#2}
\providecommand{\BIBentrySTDinterwordspacing}{\spaceskip=0pt\relax}
\providecommand{\BIBentryALTinterwordstretchfactor}{4}
\providecommand{\BIBentryALTinterwordspacing}{\spaceskip=\fontdimen2\font plus
\BIBentryALTinterwordstretchfactor\fontdimen3\font minus \fontdimen4\font\relax}
\providecommand{\BIBforeignlanguage}[2]{{%
\expandafter\ifx\csname l@#1\endcsname\relax
\typeout{** WARNING: IEEEtran.bst: No hyphenation pattern has been}%
\typeout{** loaded for the language `#1'. Using the pattern for}%
\typeout{** the default language instead.}%
\else
\language=\csname l@#1\endcsname
\fi
#2}}
\providecommand{\BIBdecl}{\relax}
\BIBdecl

\bibitem{kolmogorov_vityani}
M.~Li and P.~M. Vitnyi, \emph{An Introduction to Kolmogorov Complexity and Its Applications}, 3rd~ed.\hskip 1em plus 0.5em minus 0.4em\relax Springer Publishing Company, Incorporated, 2008.

\bibitem{tishby2000informationbottleneckmethod}
\BIBentryALTinterwordspacing
N.~Tishby, F.~C. Pereira, and W.~Bialek, ``The information bottleneck method,'' 2000. [Online]. Available: \url{https://arxiv.org/abs/physics/0004057}
\BIBentrySTDinterwordspacing

\bibitem{NIPS2003_gaussian_bottleneck}
G.~Chechik, A.~Globerson, N.~Tishby, and Y.~Weiss, ``Information bottleneck for gaussian variables,'' in \emph{Advances in Neural Information Processing Systems}, S.~Thrun, L.~Saul, and B.~Sch\"{o}lkopf, Eds., vol.~16.\hskip 1em plus 0.5em minus 0.4em\relax MIT Press, 2003.

\bibitem{theodorouGeneralizedPathIntegral}
E.~A. Theodorou, J.~Buchli, S.~Schaal, and B.~Org, ``A {{Generalized Path Integral Control Approach}} to {{Reinforcement Learning}},'' p.~45.

\bibitem{williamsInformationTheoreticMPC2017}
G.~Williams, N.~Wagener, B.~Goldfain, P.~Drews, J.~M. Rehg, B.~Boots, and E.~A. Theodorou, ``Information theoretic {{MPC}} for model-based reinforcement learning,'' in \emph{2017 {{IEEE International Conference}} on {{Robotics}} and {{Automation}} ({{ICRA}})}.\hskip 1em plus 0.5em minus 0.4em\relax {Singapore}: {IEEE}, May 2017, pp. 1714--1721.

\bibitem{rate_vs_cost_2017}
V.~Kostina and B.~Hassibi, ``Rate-cost tradeoffs in control,'' in \emph{2016 54th Annual Allerton Conference on Communication, Control, and Computing (Allerton)}, 2016, pp. 1157--1164.

\bibitem{control_comm_constraints_2004}
S.~Tatikonda and S.~Mitter, ``Control under communication constraints,'' \emph{IEEE Transactions on Automatic Control}, vol.~49, no.~7, pp. 1056--1068, 2004.

\bibitem{LQG_comm_constrainted_1998}
S.~Tatikonda, A.~Sahai, and S.~Mitter, ``Control of lqg systems under communication constraints,'' in \emph{Proceedings of the 37th IEEE Conference on Decision and Control (Cat. No.98CH36171)}, vol.~1, 1998, pp. 1165--1170 vol.1.

\bibitem{Borkar1997}
\BIBentryALTinterwordspacing
V.~S. Borkar and S.~K. Mitter, \emph{LQG Control with Communication Constraints}.\hskip 1em plus 0.5em minus 0.4em\relax Boston, MA: Springer US, 1997, pp. 365--373. [Online]. Available: \url{https://doi.org/10.1007/978-1-4615-6281-8_21}
\BIBentrySTDinterwordspacing

\bibitem{Tishby_past_future_bottleneck_2015}
N.~Amir, S.~Tiomkin, and N.~Tishby, ``Past-future information bottleneck for linear feedback systems,'' in \emph{2015 54th IEEE Conference on Decision and Control (CDC)}, 2015, pp. 5737--5742.

\bibitem{Tishby_min_info_LQG_2016}
R.~Fox and N.~Tishby, ``Minimum-information lqg control part i: Memoryless controllers,'' in \emph{2016 IEEE 55th Conference on Decision and Control (CDC)}, 2016, pp. 5610--5616.

\bibitem{separation_principle_bottleneck_2017}
A.~Achille and S.~Soatto, ``A separation principle for control in the age of deep learning,'' \emph{Annual Review of Control, Robotics, and Autonomous Systems}, vol.~1, no. Volume 1, 2018, pp. 287--307, 2018.

\bibitem{task_driven_estimation_control_ICRA_2018}
V.~Pacelli and A.~Majumdar, ``Task-driven estimation and control via information bottlenecks,'' in \emph{2019 International Conference on Robotics and Automation (ICRA)}.\hskip 1em plus 0.5em minus 0.4em\relax IEEE Press, 2019, p. 2061–2067.

\bibitem{task_driven_control_bottleneck_RSS_2020}
------, ``Learning task-driven control policies via information bottlenecks,'' \emph{Robotics: Science and Systems (RSS)}.

\bibitem{nonlinear_bottleneck}
A.~Kolchinsky, B.~D. Tracey, and D.~H. Wolpert, ``Nonlinear information bottleneck,'' \emph{Entropy}, vol.~21, no.~12, 2019.

\end{thebibliography}
\bibliographystyle{IEEEtran}

\end{document}